\documentclass[11pt]{article}
\usepackage{amsfonts}
\usepackage{pstricks}
\usepackage{pst-node}
\usepackage{epsfig}
\usepackage{epsfig}
\usepackage[latin5]{inputenc}
\usepackage[T1]{fontenc}

\begin{document}
	\def\ba{\begin{eqnarray}}
	\def\ea{\end{eqnarray}}
	\def\w{\wedge}
	\def\d{\mbox{d}}
	\def\D{\mbox{D}}

\begin{titlepage}	
\title{Gravitational Plane Waves in a Non-Riemannian Description of Brans-Dicke Gravity}
\author{Tekin Dereli\footnote{tdereli@ku.edu.tr},     
Yorgo \c{S}eniko\u{g}lu\footnote{ysenikoglu@ku.edu.tr}\\
{\small  Department of Physics, Ko\c{c} University, 34450 Sar{\i}yer-\.{I}stanbul, Turkey}  }

\date{10 October 2019}
 \maketitle	
 	

\begin{abstract}
\noindent 
The gravitational field equations of Brans-Dicke theory are given in a 4-dimensional non-Riemannian space-time with torsion in the language of exterior differential forms. 
A class of pp-wave metrics together with the Brans-Dicke scalar field are used to derive  the autoparallel equations of motion for non-spinning test masses. 
These are compared with the geodesic equations of motion and the differences are pointed out.  
The effects of the gradient of the Brans-Dicke  scalar on the geodesic deviation equations in this non-Riemannian setting are also discussed. 
 \end{abstract}
 

\noindent PACS numbers: 04.50.Kd, 04.30.-w, 95.30.sf

\end{titlepage}

\newpage

 \section{Introduction}
Einstein's general relativity is a relativistic field theory of gravitation based on a curved pseudo-Riemannian geometry that is determined solely by the metric of spacetime.  As an alternative to the general theory of relativity, Brans and Dicke theory of gravity\cite{B1, B2, B3}, among many scalar-tensor theories of gravity, describes a modification of Einstein's original formulation. In the classical Einstein theory, the metric tensor, specifically geometry, amounts to gravity. Einstein asserted that his simple metric hypothesis would lead to the situation that the inertial mass of a particle would depend on the gravitational impacts of the remainder of the universe. Nonetheless, it has been called to attention by Brans that this was a momentary coordinate effect. Looking for an approach to incorporate Mach's Principle, Brans and Dicke were directed to conceive the presence of a scalar field $\phi$ that cannot be "scaled away". Current astrophysical observations and low energy string theories are filled with scalar fields and they are proficient in explaining the large scale structure of spacetime and subatomic physics. So it is rational to consider them also at large scales. 

Brans-Dicke theory of gravity incorporates the geodesic postulate; assuming gravity is a property of the spacetime geometry, the motion of a test mass would be determined by the geodesics with the gravitational effects of all particles embedded in the metric and the associated Levi-Civita connection. The equivalence principle is then ensured with the independence in $\phi$ of the matter Lagrangian. The theory comprises the use of a pseudo-Riemannian spacetime, which leads to the field equations by second order variations. The string unification of gravity and all quantum forces make scientists wonder on the property and practicality of the theory being described, by geometry only, on all scales. There are several indications that a non-Riemannian description of spacetime may provide a refined way to characterise gravitational fields\cite{B4,B5,B6}. A natural observation can be made with a first order variation of the Lagrangian in the Palatini formalism, which implies that the spacetime connection should be relaxed to admit a non-vanishing torsion. The field equations obtained by the first order (Palatini) formalism are equivalent to those obtained by the second order formalism, up to a shift in the coupling constant \cite{B7}.

In fact, the motion of a non-spinning  test mass under the influence of gravity in a pseudo-Riemannian spacetime is a geodesic with the gravitational effects embedded in the connection. In a non-Riemannian geometry, the natural generalisation is that non-spinning test masses and light rays moving under the influence of gravity alone should follow auto-parallels of the connection with torsion \cite{B8}, determined dynamically in terms of the gradient of the scalar field. In the presence of gauge fields or other matter, one has to anticipate that matter fields will modify the geometry of spacetime and if spin is considered then the torsion will be obtained by differential equations rather than algebraic ones \cite{B9,B10}. In this article, we will be interested in the effect of the parallely propagating gravitational plane wave space-time geometry with torsion - proportional to the gradient of the Brans-Dicke scalar field - on non-spinning test masses. In particular, we are motivated on exactly how the autoparallels of the connection with torsion differ from the geodesic equations of motion and how the geodesic deviation equation changes in presence of the scalar field.

The article is organized as follows: in Section 2, we present the Brans-Dicke theory of gravity with the gravitational field equations and the equations of motion of a non-spinning test mass in a non-Riemannian setting in the coordinate independent language of differential forms. Section 3 details the plane fronted gravitational waves in Rosen coordinates. We derive the first integrals of motion and integrate them to obtain the explicit solution for the coordinates as functions of proper time. Geodesic and auto-parallel equations of motion are explicitly given and compared. The geodesic deviation equations are also discussed. Section 4 is devoted to concluding remarks. 
\bigskip

\section{Brans-Dicke Gravity}

\subsection{Gravitational Field Equations}

\noindent 
Let $M$ be a 4-dimensional spacetime manifold. In order to obtain the field equations of the Brans-Dicke theory of gravitation, we start from an action $I=\int_{M} \mathcal{L}$ and use the variational principle with the action density 4-form

\begin{equation}
 {\mathcal{L}} = \frac{\alpha^2}{2}R_{ab} \wedge *(e^a \wedge e^b) -\frac{c}{2} d\alpha \wedge *d\alpha ,
  \end{equation}
Here $c$ is a real coupling constant, the space-time metric is $g=\eta_{ab}e^{a} \otimes e^b$ with $\eta_{ab}=diag(-+++)$ where  $\{e^a\}$ are the co-frame 1-forms. ${}^*$ denotes 
the Hodge map in regard to the space-time orientation $*1 = e^0 \wedge e^1 \wedge e^2 \wedge e^3$. $\phi=\alpha^2$ denotes the Brans-Dicke scalar field. $\{{\omega}^{a}_{\;\;b}\}$ are the connection 1-forms that satisfy the Cartan structure equations
  \begin{equation}
  de^a + \omega^{a}_{\;\;b}  \wedge e^b = T^a
  \end{equation}
  with the torsion 2-forms $T^a$ and  
  \begin{equation}
 d \omega^{a}_{\;\;b} + \omega^{a}_{\;\;c}  \wedge \omega^{c}_{\;\;b} =R^{a}_{\;\;b}   
  \end{equation}
with the curvature 2-forms $R^{a}_{\;\;b}$ of spacetime. 

Independent variations of the action with respect to $e^a$, $\omega^{a}_{\;\;b}$ and
  the scalar field $\alpha$ lead to the coupled field equations $(c \neq 0)$:
  \begin{eqnarray}
 -\frac{\alpha^2}{2} R^{bc} \wedge *(e_a \wedge e_b \wedge e_c) &=& c\hspace{1mm}\tau_a[\alpha], \nonumber \\ 
  T^a &=& e^a \wedge \frac{d \alpha}{\alpha} ,   \nonumber \\ 
    c d*d\alpha^2 &=& 0; 
 \end{eqnarray}
  where
  \ba
  \tau_a[\alpha]=\frac{1}{2} \left (  \iota_a d\alpha *d\alpha + d\alpha\wedge \iota_a *d\alpha  \right ) \equiv T_{ab} *e^b
  \ea
 correspond to the energy-momentum 3-forms of the scalar field and $\iota_a$ stands for the interior products that satisfy 
$\iota_a e^b = \delta^{b}_{\;\; a}$.  
  The equations match up to the classical Brans-Dicke equations on the condition that we recognize the Brans-Dicke parameter $\omega$ as 
  \ba
  c=4\omega+6.
  \ea
 
A further simplification  of the Einstein field equations occurs as follows. Consider the field equations 
\ba
 -\frac{\alpha^2}{2} R^{bc} \wedge *(e_a \wedge e_b \wedge e_c) = \frac{c}{2} \left (  \iota_a d\alpha *d\alpha + d\alpha\wedge \iota_a *d\alpha  \right )
 \ea
 and take its trace  by considering the exterior product with $e^a$ from the left. From the definitions we get
 $$
 \alpha^2 {\mathcal R} *1 = c d\alpha \wedge *d\alpha.
 $$
 Therefore  
 \ba
 \alpha^2 {\mathcal R} *e_a = c \iota_{a} (d\alpha \wedge *d\alpha)=c (\iota_{a} d\alpha) *d\alpha - c d\alpha \wedge \iota_{a}  *d\alpha.
 \ea
Since
 \ba
  -\frac{\alpha^2}{2} R^{bc} \wedge *(e_a \wedge e_b \wedge e_c) = *Ric_a -\frac{1}{2}{\mathcal R} *e_a,
 \ea
 substituting for the curvature scalar in the line above and taking it to the right hand side one can show that the Einstein field equations can be given in a simpler 
 but equivalent form as
 \ba
 \alpha^2 Ric_a = c (\iota_a d\alpha) d\alpha.
 \ea
 
 \bigskip

  \subsection{Equations of Motion of a Non-spinning Test Mass}
 
 \noindent  In order to discuss the equations of motion of a non-spinning test mass we find it convenient to 
 switch to an equivalent description of the spacetime geometry in terms of a metric-compatible, type-preserving covariant derivative  
 $\nabla_X$ with respect to an arbitrary vector field $X$.  
Then the Cartan structure equations read
\ba
\nabla_X Y -\nabla_Y X - [X,Y] = T(X,Y)
\ea
that determines the type-(2,1) torsion tensor T of the connection and 
\ba
\nabla_X \nabla_Y Z -\nabla_Y \nabla_X Z -  \nabla_{[X. Y]} Z = {\mathbf R}_{Z}(X,Y)   
\ea
that determines through the definition 
\ba
\beta ( {\mathbf R}_{Z}(X,Y) ) = Riem(X,Y,Z,\beta) 
\ea
the type-(3,1) Riemann curvature tensor $Riem$ of the connection.

Brans and Dicke postulated independently of their field equations that the non-spinning test masses should follow geodesic equations of motion given by
 \ba
{\hat{\nabla}}_{\dot{C}}\dot{C}= 0,  
\ea
where $\hat{\nabla}$ is the unique Levi-Civita connection of $g$.
Here $C:[0,1] \rightarrow M$ is a curve in space-time. 
$\dot{C} = \frac{d}{d\tau} C $ denotes the unit time-like tangent vector field along the curve $C$, parametrised by the proper time $\tau$ so that $g(\dot{C},\dot{C})=-1$  
(the speed of light $c=1$).
On the other hand  here
the auto-parallel equations of motion relative to the non-Riemannian connection $\nabla$ will be assumed: 
\ba
{\nabla}_{\dot{C}}\dot{C}= 0.  
\ea
Auto-parallel curves in a non-Riemannian space-time differ in general from the geodesic curves.
Given the torsion 2-forms in Brans-Dicke gravity as above, one may  show that the auto-parallel equations of motion of a test mass 
can be written in the following way\cite{B9}:
\ba
 {\widetilde{{{\hat{\nabla}}}_{\dot{C}} \dot{C}}} = \iota_{\dot{C}}( \widetilde{\dot{C}} \wedge \frac{d\alpha}{\alpha}).  
\ea


\noindent Let us now consider a normal vector field $X$ along an auto-parallel $C$ such that $[X,\dot{C}]=0$. Under these assumptions the Cartan structure equations reduce to
\ba
\nabla_{\dot{C}}X - \nabla_{X} \dot{C} = T(\dot{C},X),
\ea
 and 
 \ba
 \nabla_{\dot{C}}(\nabla_{X} \dot{C}) -  \nabla_{X}(\nabla_{\dot{C}} \dot{C}) = {\mathbf{R}}_{\dot{C}}(\dot{C},X),
 \ea
 respectively. If the auto-parallel equations of motion  and the first set of Cartan equations are used to simplify the second set of Cartan equations, we arrive at the geodesic deviation equation
\ba
 \nabla_{\dot{C}}(\nabla_{\dot{C}} X)  = -{\mathbf{R}}_{\dot{C}}(X,\dot{C}) - \nabla_{\dot{C}} T(X,\dot{C})
\ea
which determines the normal acceleration of a non-spinning test mass moving along an auto-parallel curve in terms of the space-time curvature and torsion. 

\section{Gravitational Waves}

Gravitational waves for the class of the Poincar\'{e} gravity models with the most general Lagrangian which includes all possible linear and quadratic invariants of the torsion and the curvature are derived in \cite{B10-a}. A theoretical basis for the determination of the gravitational field is given in the study of the geodesic equation \cite{B10-b}. The geodesic deviation idea can be also extended to calculate approximate orbits of point masses in gravitational fields which is of practical applicability to the problem of the emission of gravitational radiation\cite{B10-c}. Puetzfeld and Obukhov \cite{B10-d,B10-e} have worked on spacetimes with torsion detailing the dynamics of two adjacent worldlines; in their case, they developed Synge's world function. They distinctly illustrate how the deviation equation can be used to measure the curvature of spacetime and thereby the gravitational field. The Brans-Dicke field equations can be formulated in a pseudo-Riemannian or non-Riemannian spacetime. If we consider them in a non-Riemannian spacetime, a possibility arises: the autoparallel equations of motion can be postulated . Brans and Dicke, on the other hand, postulated the geodesic hypothesis. These alternatives differ in general and we want to demonstrate the differences on a gravitational plane wave spacetime. In this regard, we will be developing in what follows  the plane fronted gravitational waves in a non-Riemannian setting where the spacetime torsion is determined by the gradient of the Brans-Dicke scalar.
\subsection{Rosen Coordinates}
Let us recall  the plane fronted gravitational wave metric in Rosen coordinates given by
\ba
g=2 \d u \d v + \frac{\d x^2}{f(u)^2} + \frac{\d y^2}{h(u)^2}  \label{Rosen},
\ea
which describes a gravitational wave that propagates along the negative $z$-axis whose wave front coinciding with the $xy$-plane and admiting non-twisting parallel rays.
It is possible to write the null coordinates as 
 \ba
u= \frac{z+t}{\sqrt{2}} \; , \quad v = \frac{z-t}{\sqrt{2}}. 
\ea     
We further take a scalar field 
\ba
\alpha&=& \alpha(u).
\ea
We note that the scalar field equation to the system is identically satisfied.

Through the coordinate transformation
\ba
U=u, \quad V=v + \frac{x^2}{2}\frac{f'}{f^3} + \frac{y^2}{2}\frac{h'}{h^3}, \quad X=\frac{x}{f}, \quad Y=\frac{y}{h},
\ea
where ${}^\prime$ denotes the derivative  $\frac{\partial}{\partial u}$, the common description of the family of pp-waves can be given in Kerr-Schild form in Brinkmann coordinates (U,V,X,Y) as follows\cite{B11,B11a}:
\ba
g=2\d U \d V + \d X^2 +\d Y^2 +2H(U,X,Y) \d U^2.  \label{brinkman}
\ea
Nevertheless the Rosen coordinates have certain advantages over the Brinkmann coordinates. First of the all, the equations of motion can be fully integrated. Another instance, both $u$ and $v$ being null coordinates, if a head-on collision of two pp-waves is assumed at some point, they can be represented in the same picture\cite{B12}. 

Working out the expressions for the curvature, torsion and  the  scalar field stress-energy-momentum tensors with $(\ref{Rosen})$, the Einstein field equations to be solved 
reduce to the following second order differential equation: 
\ba
\frac{f^{\prime\prime}}{f} -2 (\frac{f^\prime}{f})^2 + \frac{h^{\prime\prime}}{h} -2(\frac{h^\prime}{h})^2 =2 \frac{\alpha^{\prime\prime}}{\alpha}+(c-4) (\frac{\alpha^\prime}{\alpha})^2. \label{eq:FieldEq}
\ea

\noindent The auto-parallel equations of motion read 
\ba
\ddot{u}+\frac{\dot{\alpha}}{\alpha}\;\dot{u}=0, \nonumber\\
\ddot{v} - \frac{\dot{\alpha}}{\alpha}\dot{v} + (\frac{\dot{f}}{f^3}-\frac{\dot{\alpha}}{\alpha f^2})\frac{\dot{x}^2}{\dot{u}}+ (\frac{\dot{h}}{h^3}-
\frac{\dot{\alpha}}{\alpha h^2})\frac{\dot{y}^2}{\dot{u}}=0, \nonumber\\
\ddot{x}+(\frac{\dot{\alpha}}{\alpha}-2\frac{\dot{f}}{f})\dot{x}=0,\quad
\ddot{y}+(\frac{\dot{\alpha}}{\alpha}-2\frac{\dot{h}}{h})\dot{y}=0.\label{eq:autoparallel}
\ea

\noindent The first equation can be integrated once immediately:
\ba
\dot{u}=\frac{1}{\alpha(u)}\frac{p_u}{m}.
\ea
Using this, the remaining equations can also be integrated once as follows:
\ba
\dot{v}=-\frac{f^2}{2p_um}\frac{p_x^2}{\alpha}-\frac{h^2}{2p_um}\frac{p_y^2}{\alpha}-\frac{m\alpha}{2p_u}, \quad
\dot{x}=\frac{p_xf^2}{\alpha m}, \quad
\dot{y}=\frac{p_yh^2}{\alpha m}.
\ea
The constants $p_u$, $p_x$ and $p_y$ are first integrals of motion. 
Hence an explicit solution for the coordinates as functions of proper time $\tau$ is determined:
\ba
\tau=\frac{m}{p_u}\int_{0}^{\frac{p_u\tau}{m\alpha}}\alpha(u)du, \nonumber\\ 
v(\tau)=v(0)-\frac{m^2}{2p_u^2}\int_{0}^{\frac{p_u\tau}{m\alpha}}\alpha(u)^2du-\frac{p_x^2}{2p_u^2}\int_{0}^{\frac{p_u\tau}{m\alpha}}f^2(u)du-\frac{p_y^2}{2p_u^2}\int_{0}^{\frac{p_u\tau}{m\alpha}}h^2(u)du, \nonumber\\ 
x(\tau)=x(0)+\frac{p_x}{p_u}\int_{0}^{\frac{p_u\tau}{m\alpha}}f^2(u)du, \quad
y(\tau)=y(0)+\frac{p_y}{p_u}\int_{0}^{\frac{p_u\tau}{m\alpha}}h^2(u)du.
\ea
\noindent
We immediately notice that with the Rosen form we have obtained explicit first integrals of motion and solutions for the coordinates as functions of proper time.
The coordinates $(t,z,x,y)$ are basically not equivalent to the measured intervals in spacetime; despite what might be expected, they are precisely the coordinates balanced for the trajectories of initially stationary, non-interacting test masses. The displacement of particles in the $x$ (or $y$) direction is actually $f^{-1}dx$ (resp. $h^{-1}dy$), which varies with $u$ i.e. $\frac{t+z}{\sqrt{2}}$ for fixed $dx$ and $dy$. Thus a gravity wave causes acceleration of a test mass perpendicular to its direction of propagation. 

By fixing a scale $\alpha=1$, we can obtain the geodesic equations of motion. An explicit solution for the coordinates as functions of proper time $\tau$ can be given \cite{B13}  neatly as 
\ba
u(\tau)=u(0)+\frac{p_u\tau}{m} , \nonumber \\
v(\tau)=v(0)-\frac{m\tau}{2p_u}-\frac{p_x^2}{2p_u^2}\int_0^\tau f(u)^2du-\frac{p_y^2}{2p_u^2}\int_0^\tau h(u)^2du ,\nonumber\\
x(\tau)=x(0)+\frac{p_x}{p_u}\int_0^\tau f(u)^2du, \quad 
y(\tau)=y(0)+\frac{p_y}{p_u}\int_0^\tau h(u)^2du.
\ea
By comparing the equations (29) and (30) above, the autoparallel and geodesic equations of motion differ. We remark that in the geodesic equations of motion $u(\tau)$ is linear in $\tau$; so all of the limits of the integrals are written from $0$ to $\tau$. The effect of the scalar field $\alpha$ is present in every autoparallel equations of motion. We note that the proper times in the geodesic and autoparallel equations of motion are scaled with respect to each other. This is a subtle point, in each framework, the proper time is defined in a different way. It is not only geodesic and autoparallel curves that are different but the presence of the scalar field $\alpha$ scales the proper time coordinate itself. We cannot think of an observation that may distinguish between these two cases. That is why looking at the effects in the geodesic deviation equations could be crucial.

\subsection{Geodesic Deviation Equations }

\noindent Suppose we are given the parametric representation of a spacetime curve $C:[0,1] \rightarrow M$ in a local  chart $ x^\mu = x^\mu (\tau)$. Then the tangent vector field    
along the curve will be given in Rosen coordinates $(u,v,x,y)$ by 
\ba
\dot{C} = \dot{u} \partial_u  +  \dot{v} \partial_v + \dot{x} \partial_x + \dot{y} \partial_y
\ea
It is going to be a unit, time-like vector field provided
\ba
g(\dot{C},\dot{C}) = 2\dot{u}\dot{v} + \frac{{\dot{x}}^2}{f(u)^2} +\frac{{\dot{y}}^2}{h(u)^2} = -1.
\ea
Here we consider  a solution curve of the autoparallel equations of motion $\nabla_{\dot{C}}\dot{C} = 0.$
A Jacobi normal vector field  $X$  satisfies  $[\dot{C},X]=0$ at all points along $C$.
One convenient choice for the Jacobi vector field would be
\ba
X = \eta^{1}(u,x,y) \partial_x  +  \eta^{2}(u,x,y) \partial_y
\ea
provided 
\ba
\dot{u}(\partial_u \eta^1)+\dot{x}(\partial_x \eta^1)+\dot{y}(\partial_y \eta^1)  = 0 = 
\dot{u}(\partial_u \eta^2)+\dot{x}(\partial_x \eta^2)+\dot{y}(\partial_y \eta^2).
\ea
The Jacobi vector field $X$ restricted to the curve $C$ satisfies a geodesic deviation equation that will be modified in a non-Riemannian setting. 
We work out explicitly the  geodesic deviation equation in general for autoparallels as follows:

\ba
0&=&\Big(\frac{f''}{f^3} - \frac{2f'^2}{f^4} -\frac{\alpha''}{\alpha f^2}+\frac{3\alpha'^2}{2\alpha^2 f^2}+\frac{\alpha' f'}{2\alpha f^3}\Big)(\dot{u}\dot{x})\eta^1\nonumber\\
&&+\Big(\frac{h''}{h^3} - \frac{2h'^2}{h^4} -\frac{\alpha''}{\alpha h^2}+\frac{3\alpha'^2}{2\alpha^2 h^2}+\frac{\alpha' h'}{2\alpha h^3}\Big)(\dot{u}\dot{y})\eta^2,\nonumber\\
\ddot{\eta}^1&=&\frac{\dot{\alpha}}{2\alpha}\dot{\eta}^1+\Bigg[\Big(-\frac{f''}{f} + \frac{2f'^2}{f^2}-\frac{\alpha' f'}{2\alpha f} + \frac{1}{2}(\frac{\alpha'}{\alpha})'\Big)\dot{u}^2+\Big(\frac{{\alpha'}}{2\alpha}\Big)\ddot{u}\Bigg]\eta^1\nonumber\\
\ddot{\eta}^2&=&\frac{\dot{\alpha}}{2\alpha}\dot{\eta}^2 + \Bigg[\Big(\frac{h''}{h} - \frac{2h'^2}{h^2}-\frac{\alpha' h'}{2\alpha h}+\frac{1}{2}(\frac{\alpha'}{\alpha})'\Big)\dot{u}^2+\Big(\frac{{\alpha'}}{2\alpha}\Big)\ddot{u}\Bigg]\eta^2.  \label{eq:Geo-Dev}
\ea
The special case $\alpha=1$ gives us back the commonly known expressions:
\ba
0&=&\Bigg[\Big(\frac{f''}{f^3} - \frac{2f'^2}{f^4} \Big)(\dot{u}\dot{x})\Bigg]\eta^1+\Bigg[\Big(\frac{h''}{h^3} - \frac{2h'^2}{h^4}\Big)(\dot{u}\dot{y})\Bigg]\eta^2,\nonumber\\
\ddot{\eta}^1&=&\Bigg[\Big(-\frac{f''}{f} + \frac{2f'^2}{f^2}\Big)\dot{u}^2\Bigg]\eta^1, \quad
\ddot{\eta}^2= \Bigg[\Big(\frac{h''}{h} - \frac{2h'^2}{h^2}\Big)\dot{u}^2\Bigg]\eta^2.\label{eq:Geo-Dev}
\ea
We note that similar considerations can be found in \cite{B14}, that detail the motion of test particles with an arbitrary curvature-matter coupling. 
The geodesic deviation equations in general from the point of view of Hamiltonian dynamics and symmetries of the dynamics are discussed in \cite{B15}. Here, instead of giving a detailed analysis of integrability of the system (35), we wish to consider the geodesic deviation equations in some simple cases and point out possible effects of the scalar field $\alpha$.
Let us imagine that we have a gravitational wave antenna consisting of a number of non-interacting, non-spinning test masses that are distributed uniformly on a circle in the $xy$-plane. They remain in equilibrium when there is no gravitational wave since they are non-interacting. If a gravitational plane wave hits the antenna orthogonally along its symmetry axis and assuming all the coefficients in (35) are slowly varying and negative, the geodesic deviation equations will imply that the circle would oscillate periodically in the transverse direction. But the geodesic deviation equations for autoparallels (35) show that there may be a damping on such oscillations, explicitly produced by the terms that involve $\dot{\eta_1}$ and $\dot{\eta_2}$. 


\section{Concluding Remarks}
In this article, we study mainly the plane fronted gravitational waves in Brans-Dicke gravity in a non-Riemannian setting. The spacetime geometry is relaxed to admit torsion, that depends linearly on the gradient of the scalar field. We discuss explicit solutions whereas the geodesic and autoparallel equations of motion of a non-spinning test mass  
are integrated to the end in Rosen coordinates. 
In particular, we note that the differences between autoparallel equations of motion in general in the presence of the gradient of Brans-Dicke scalar $\alpha$ and the geodesics equations of motion can be  recognised  by comparison.

If an explicit solution in terms of metric functions $f(u), h(u)$ and $\alpha$ is given; one may evaluate the first integrals of motion. The constants appearing in (28) are not all independent as they also satisfy the orbit equation (32). 
Then substituting these in the geodesic and/or autoparallel equations of motion, explicit solutions for the coordinates as functions of proper time $\tau$ are found by integrating the first integrals once more. Furthermore, the definition of proper time in both cases differ from each other. By the insertion of these solutions in the geodesic deviation equation (19) related with the world-lines of two neighbouring, non-interacting spinless test masses, the effects of the spacetime curvature and torsion can be derived. Thus a comparison of geodesic and auto-parallel cases would follow. If it were possible to send a gravitational plane wave that propagates orthogonally to the transverse plane, we could have detected the oscillatory motion of the point masses and decide whether they move along geodesic or autoparallel curves. This does give us the opportunity to observe the difference experimentally, but the observation of a damping effect is critical as it would definitely imply that the test masses follow autoparallels.  

Brans and Dicke have originally presented their geodesic postulate in a subtle way by assuming that the matter Lagrangian should be independent of their scalar field $\phi$. It is well known that the Brans-Dicke field equations may be related with the Einstein massless scalar field equations by a suitable scaling of the metric and a scalar field redefinition. It is natural to assume in the Einstein picture to consider the corresponding geodesic equations of motion. However, with the same field redefinitions that relate these two pictures, the geodesic equations of motion in the Einstein massless scalar field theory are transformed to the autoparallel equations of motion in Brans-Dicke theory.

\section{Acknowledgement}
Y.\c{S}. is grateful to Ko\c{c} University for its hospitality and partial support.

\bibliography{references}
\bibliographystyle{iopart-num}

\end{document}